\documentclass[reprint,twocolumn,notitlepage,preprintnumbers,nofootinbib,amsmath,amssymb,aps,prd,floatfix,superscriptaddress,longbibliography,svgnames]{revtex4-1}

\usepackage{amsmath}
\usepackage{amsfonts}
\usepackage{amssymb}
\usepackage{graphicx}
\usepackage{bm}
\usepackage{subfigure}
\usepackage{url}
\usepackage[hyperindex]{hyperref}
\usepackage{color}
\usepackage[ddmmyy,24hr]{datetime}
\usepackage{bigdelim}
\usepackage{booktabs}
\usepackage{dcolumn}
\usepackage{multirow}
\usepackage{subfigure}
\usepackage{physics}
\usepackage{cancel}
\usepackage{stackrel}
\usepackage{paralist}
\usepackage{xspace}
\usepackage{slashed}
\usepackage{cancel}
\usepackage{todonotes}
\usepackage{enumerate}
\usepackage{float}
\usepackage{fullpage}
\usepackage{ulem}
\usepackage{wasysym}
\usepackage{lineno}

\usepackage{comment}

\begin{document}
\title{Simultaneous Extraction of the Weak Radius and the Weak Mixing Angle \\ from Parity-Violating Electron Scattering on $^{12}\mathrm{C}$}

\author{M. Cadeddu}
\email{matteo.cadeddu@ca.infn.it}
\affiliation{Istituto Nazionale di Fisica Nucleare (INFN), Sezione di Cagliari,
	Complesso Universitario di Monserrato - S.P. per Sestu Km 0.700,
	09042 Monserrato (Cagliari), Italy}
\author{N. Cargioli}
\email{nicola.cargioli@ca.infn.it}
\affiliation{Dipartimento di Fisica, Universit\`{a} degli Studi di Cagliari,
	Complesso Universitario di Monserrato - S.P. per Sestu Km 0.700,
	09042 Monserrato (Cagliari), Italy}
\affiliation{Istituto Nazionale di Fisica Nucleare (INFN), Sezione di Cagliari,
	Complesso Universitario di Monserrato - S.P. per Sestu Km 0.700,
	09042 Monserrato (Cagliari), Italy}
 \author{J. Erler}
 \email{erler@uni-mainz.de}
 \affiliation{PRISMA$^+$ Cluster of Excellence, Institute for Nuclear Physics, Johannes Gutenberg-University, 55099 Mainz, Germany}

\author{M. Gorchtein}
\email{gorshtey@uni-mainz.de}
\affiliation{PRISMA$^+$ Cluster of Excellence, Institute for Nuclear Physics, Johannes Gutenberg-University, 55099 Mainz, Germany}

\author{J. Piekarewicz}
\email{jpiekarewicz@fsu.edu}
\affiliation{Department of Physics, Florida State University, 
             Tallahassee, FL 32306-4350, USA}

\author{Xavier Roca-Maza}
\email{xavier.roca.maza@fqa.ub.edu}
\affiliation{INFN, Sezione di Milano, 20133 Milano, Italy}
\affiliation{Dipartimento di Fisica ``Aldo Pontremoli'', Universit\`a degli Studi di Milano, 20133 Milano, Italy}
\affiliation{Departament de F\'isica Qu\`antica i Astrof\'isica, Mart\'i i Franqu\'es, 1, 08028 Barcelona, Spain}
\affiliation{Institut de Ci\`encies del Cosmos, Universitat de Barcelona, Mart\'i i Franqu\'es, 1, 08028 Barcelona, Spain}

\author{H. Spiesberger}
\email{spiesber@uni-mainz.de}
\affiliation{PRISMA$^+$ Cluster of Excellence, Institute for Nuclear Physics, Johannes Gutenberg-University, 55099 Mainz, Germany}

\begin{abstract}
We study the impact of nuclear structure uncertainties on a measurement of the weak charge 
of $^{12}\mathrm{C}$ at the future MESA facility in Mainz. Information from a large variety of 
nuclear models, accurately calibrated to the ground-state properties of selected nuclei, 
suggest that a $0.3$\,\% precision measurement of the parity-violating asymmetry at forward 
angles will not be compromised by nuclear structure effects, thereby allowing a world-leading determination of the 
weak charge of $^{12}\mathrm{C}$. Furthermore, we show that a combination of measurements of the 
parity-violating asymmetry at forward and backward angles for the same electron beam energy can be used to extract information 
on the nuclear weak charge distribution. We conclude that a $0.34$\% precision on the weak 
radius of $^{12}\mathrm{C}$ may be achieved by performing a $3$\% precision measurement of the parity-violating asymmetry at backward angles. 
\end{abstract}

\maketitle 
\section{Introduction}
The weak mixing angle $\theta_\mathrm{W}$ is the fundamental parameter of the electroweak sector of the standard model (SM), which describes the mixing between the SU(2) and U(1) gauge  fields. Over the years, many measurements of the weak mixing angle have been performed to test the SM and to search for possible hints of beyond the SM effects.
Its sine squared $\sin^2\theta_\mathrm{W}$ enters the neutral current vector charge of SM fermions and can be accessed through a variety of processes at very different energy scales: from the ``Z-pole" measurements performed at colliders \cite{ALEPH:2005ab,CDF:2018cnj,LHCb:2015jyu,Andari:2019tfc,CMS:2024ony}, to deep inelastic scattering with electrons \cite{PVDIS:2014cmd,Prescott:1979dh}, and neutrinos \cite{NuTeV:2001whx}, to parity violation in atoms \cite{Wood:1997zq,Guena:2004sq}, and parity-violating electron scattering (PVES) off protons \cite{Qweak:2018tjf} and electrons \cite{SLACE158:2005uay}.
At low energies, also the coherent elastic neutrino-nucleus scattering process can provide a determination of the weak mixing angle~\cite{AtzoriCorona:2023ktl}. Although the current precision is still much worse than the other available probes, it provides complementary information to atomic parity violation and PVES measurements with nuclei~\cite{AtzoriCorona:2024vhj}.
A set of precise measurements at different energy scales are needed to test the running of the weak mixing angle, which arises from loop-level contributions~\cite{Erler:2004in,Erler:2017knj}. Our current theoretical knowledge of this running is much more precise than the corresponding experimental accuracy, especially at low energies. Future, more precise low-energy measurements are thus necessary to further test the SM and to disentangle beyond the SM scenarios. There are efforts to exploit atomic parity violation to measure the weak charge of heavy nuclei at the per mille precision \cite{Antypas:2018mxf,Zhang:2016czr}. On the PVES side, the Qweak experiment has reached a 0.5\% accuracy \cite{Qweak:2018tjf,Qweak:2021ijt}, and the upcoming  P2@MESA \cite{Becker:2018ggl} and the MOLLER \cite{MOLLER:2014iki} experiments aim at improving this precision by factors of 4 and 6, respectively.
\\
PVES on nuclei is not only useful for testing the SM, but it also allows one to gain information on the structure of the target nucleus which {has been traditionally accessible only via strongly interacting probes that could carry significant systematic unceratainties of the experimental analysis}. In recent years, PVES off neutron-rich nuclei has been used to determine a crucial nuclear property known as the neutron skin thickness, defined as the difference {between the root mean square radii of neutron and proton nuclear distributions. Intimately connected to this non-observable quantity is the so-called weak skin, defined as the difference between root mean square radii of the weak and the charge nuclear distributions. This quantity may indeed be directly measured in the experiment. A strong motivation to study these quantities is to constrain the equation of state (EOS) of neutron-rich matter \cite{Horowitz:2000xj,Reed:2021nqk}, which is central to our understanding of the structure of neutron stars \cite{Horowitz:2000xj,Thiel:2019tkm}. 
\\
Challenging experimental programs to extract the neutron skin via PVES were carried out successfully at Jefferson Lab. In particular, the 
PREX \cite{Abrahamyan:2012gp,PREX:2021umo} and CREX \cite{CREX:2022kgg} Collaborations measured the neutron skins of $^{208}$Pb and $^{48}$Ca, respectively. Recently, also the Qweak Collaboration released a first neutron skin measurement from PVES on $^{27}\mathrm{Al}$ \cite{Qweak:2021ijt}. 
The interpretation of these experiments in terms of the neutron skin has been made possible thanks to a considerable input from the nuclear and particle theory communities.
\\
The interplay between the weak mixing angle and the neutron skin for a PVES measurement was already studied in Ref.~\cite{Corona:2021yfd} for the case of lead, while in a previous work~\cite{Koshchii:2020qkr} some of the present authors addressed the feasibility of an intriguingly precise simultaneous extraction of the weak charge and the neutron skin from a PVES experiment off a light nucleus, such as $^{12}$C. Such an approach is highly beneficial and complementary to the coherent elastic neutrino nucleus scattering (CE$\nu$NS) campaign that is sensitive to the same observables\,\cite{Akimov:2019rhz,Yang:2019pbx,AtzoriCorona:2023ktl,Cadeddu:2021ijh,AtzoriCorona:2024vhj}.
\\
In the present study, we build upon the latter work for $^{12}$C using a more advanced treatment of the information provided by a broader set of theoretical models. In particular, we include in our new study a collection of models using non-relativistic \cite{KDE, NRAPR, QMC, SKX, SIII, SGII, SKM*, SKP, SKI, SLY, SKO, SAMI, SAMIJ, SK255} and covariant \cite{Chen:2014sca,Chen:2014mza,Fattoyev:2013yaa} energy density functionals.

\section{Parity violating electron scattering}
The experimental observable of a PVES measurement is the parity-violating asymmetry, $A^{\rm PV}$, which is defined as the difference between the differential cross sections for elastic scattering of longitudinally polarized electrons off a spin-zero target, 
\begin{equation}
    A^{\rm PV}=\dfrac{\sigma_R-\sigma_L}{\sigma_R+\sigma_L},
\end{equation}
where $\sigma_{R (L)}$ represents the cross sections with right(left)-handed electron polarization. The asymmetry arises from the interference between two parts of the scattering amplitude: one due to the exchange of a virtual photon and the other one from exchanging a virtual $Z^0$-boson. 
Adopting momentarily the plane-wave Born (tree-level) approximation, the asymmetry takes the form
\begin{equation}
    A^{\rm PV}_{\rm PWBA}=-\dfrac{G_F Q^2}{4\sqrt{2}\pi \alpha}\dfrac{Q_W}{Z}\dfrac{F_{\rm wk}(Q^2)}{F_{\rm ch}(Q^2)},\label{ApvBorn}
\end{equation}
with $G_{F}$ the Fermi constant, $\alpha$ the fine-structure constant, and $Q^2$ the four-momentum transfer squared. Moreover, $Q_W$ denotes the nuclear weak charge, $Z$ the nuclear electric charge, while $F_{\rm wk}/F_{\rm ch}$ is the ratio of the weak and charge form factors that encode the nuclear structure information. The formulation in Eq.~\eqref{ApvBorn} is very useful as it clearly factorizes the dependence on fundamental SM parameters from nuclear effects encoded in the form factor ratio. To further underscore this aspect we define the normalization factor $A_0$ as follows:
\begin{equation}
   A_0=-\dfrac{G_F Q^2}{4\sqrt{2}\pi \alpha}\dfrac{Q_W}{Z},\label{ApvBornA0}
\end{equation}
which incorporates only kinematical factors, SM parameters, and the nuclear weak {and electric} charges. $A_0$ represents the value of the asymmetry for a point-like nucleus, {that is,} with the form factors equal to unity independent of the momentum transfer.
\\
The weak charge of a nucleus composed of $Z$ protons and $N$ neutrons is given by 
\begin{subequations}
\begin{align}
 & Q_{W}(Z,N) =ZQ_W^p+NQ_W^n, \;\;{\rm where} \\
 & Q_W^p=(1-4\sin^2\theta_W) \;\;{\rm and}\;\; Q_W^n=-1,
\end{align}
\end{subequations}
and the last equation denotes tree-level proton and neutron weak charges, respectively.
This expression highlights the dependence of the nuclear weak charge on the weak mixing angle. Including radiative corrections, the numerical value of the weak charge of $^{12}\mathrm{C}$ is $Q^{\rm SM}_W(^{12}\mathrm{C})\!=\!-5.499$ \cite{Erler:2014fqa}.
\\
The charge form factor in Eq.~\eqref{ApvBorn} has been extracted from elastic electron scattering data. The best fit to data is obtained by using the sum-of-Gaussians (SOG) parametrization of the density profile~\cite{DeVries:1987atn}. The slope of the charge form factor at zero momentum transfer is related to the nuclear charge radius. For $^{12}$C, we use the reference value obtained from scattering data, $R_{\rm ch}^{\rm SOG}\!=\!2.469(5)\,\mathrm{fm}$. 
X-ray spectroscopy of muonic atoms is known to usually give more precise values of nuclear charge radii; however, oftentimes the two methods give different values for the radius \cite{DeVries:1987atn,Angeli:2013epw}.
The case of carbon is special in this respect: not only are the respective charge radius extractions compatible, $R_{\rm ch}^{\mu}\!=\!2.4702(22)\,\mathrm{fm}$~\cite{Angeli:2013epw}, the value obtained from the electron scattering data is only slightly less precise. We opt to use the SOG radius for a more conservative error analysis. 
In turn, the weak 
form factor is only loosely constrained by hadronic experiments that involve large
and poorly understood systematic uncertainties\,\cite{Thiel:2019tkm}. 
\\
In the Born approximation, Eq.~\eqref{ApvBornA0}, Coulomb corrections which generically scale as $\pi\alpha Z$ ($\sim\!0.1$ for $^{12}$C) are negelected. Given the experimental precision goal of $0.3\%$, not only are these corrections essential, but they have to be resummed to all orders. 
This is done within the distorted wave Born approximation (DWBA) by numerically solving the Dirac equation for an electron in the presence of the Coulomb potential generated by the entire nuclear charge distribution. This requires to go beyond the Coulomb potential of a point charge. In addition, in order to provide predictions for the PV asymmetry, the Coulomb potential has to be supplemented with the weak interaction potential which depends on the nuclear weak charge distribution and acts on left- and right-handed electrons with the opposite sign\,\cite{Horowitz:1998vv}.
We note that one can go beyond the tree level in DWBA calculations by including higher-order QED corrections such as the vacuum polarization (Uehling potential) and self-energy corrections \cite{Shabaev2000}. These two contributions are expected to partially cancel each other 
\cite{Shabaev2000,Roca-Maza2017}.
We do not include these corrections in this work since they are much smaller than the uncertainties associated with the nuclear weak form factor. 
Finally, we note that for numerical calculations, we employ the DREPHA code~\cite{Drepha}. We checked its consistency with the ELSEPA package~\cite{Elsepa} which was used in Ref.~\cite{Koshchii:2020qkr}.

\section{Theoretical Framework}
Given that the main nuclear inputs to the parity-violating asymmetry are ground-state densities and their associated form factors, we start
this section by reviewing the theoretical framework underpinning our predictions. Since such a theoretical approach assumes point nucleons, we then proceed to discuss how single nucleon form factors are used to refine our predictions. We end this section by quantifying the theoretical uncertainties that affect the precision by which $\sin^2\theta_\mathrm{W}$ may be determined from a measurement of the parity violating asymmetry with ${}^{12}$C. 

\subsection{Density Functional Theory}\label{sec:DFT}
Predictions for the parity violating asymmetry for ${}^{12}$C will be made by a collection of both non-relativistic and relativistic 
(or covariant) energy density functionals\,\cite{KDE, NRAPR, QMC, BSK17, SKX, SIII, SGII, SKM*, SKP, SKI, SLY, SKO, SAMI, SAMIJ, SK255,Chen:2014sca,Chen:2014mza,Fattoyev:2013yaa}. The energy density functional (EDF) is the main building block behind Density Functional Theory (DFT), a powerful technique developed by Kohn and collaborators\,\cite{Hohenberg:1964zz,Kohn:1965} to understand the electronic structure of complex many-body systems, but since then extended to many other areas of physics. Perhaps the most remarkable result of DFT is that the exact ground-state energy of the complicated many-body system may be obtained from minimizing a suitable 
EDF that only depends on the one-body density. By doing so, DFT not only reduces drastically the complexity of the problem, but also 
benefits from incorporating physical insights into the construction of the functional. This is particularly relevant given that the DFT formalism is formulated as an existence theorem that offers no guidance on how to construct the appropriate EDF. In nuclear physics, one incorporates a myriad of physical insights into the construction of the functional and leaves the parameters of the model to be determined through a calibration procedure informed by the ground-state properties of finite nuclei. 

Inspired by the long history of Hartree-Fock theory in a variety of disciplines, Kohn and Sham replaced the explicit construction of 
the energy density functional by an equivalent system of non-interacting particles moving in a suitably generated one-body 
potential\,\cite{Kohn:1965}. It is important to underscore that the equivalent Kohn-Sham potential must be sophisticated enough 
to reproduce the exact one-body density of the interacting system. This often demands the inclusion of one-body terms that have a strong density dependence. The reformulation of the DFT problem in terms of one-particle orbitals has several advantages, first and foremost that self-consistent problems of this kind have a long history in nuclear physics. The outcome of solving the mean-field-like Kohn-Sham equations is a single-particle spectrum from which proton and neutron densities may be determined. In turn, the ground state form factors may be computed by taking the Fourier transform of the corresponding proton and neutron densities. Finally, one accounts for the finite nucleon size by incorporating nucleon form factors as described in the next section. The resulting charge and weak charge form factors computed in this manner are the sole nuclear input to the parity violating asymmetry given in 
Eq.~\eqref{ApvBorn}.

\subsection{Nucleon Form Factors}\label{sec:nucleonFF}

In order to obtain reliable theoretical predictions for the electromagnetic and weak charge densities, one must
fold the point-proton and -neutron densities obtained from the nuclear structure models with single nucleon form factors. For the electromagnetic single nucleon form factors we adopt the following simple dipole parametrization\,\cite{Kelly:2002if,Horowitz:2012we}: 
\begin{equation}
   G_{\rm E}^{\,p}(Q^{2}) = 
   \frac{G_{\rm M}^{\,p}(Q^{2})}{\mu_{p}} = 
   \frac{G_{\rm M}^{\,n}(Q^{2})}{\mu_{n}} = G_{\rm D}(Q^{2}) \;.
\label{NucleonFF}
\end{equation}
The form factors depend on the four-momentum transfer 
$Q^{2}\!\!=\!{\bf q}^{2}\!-\!\omega^{2}$
which is determined from the corresponding three-momentum transfer ${\bf q}$ and the energy loss $\omega$. 
The dipole form factor is given by
\begin{equation}
    G_{\rm D}(Q^{2}) = \left(1+\frac{Q^{2}}{12}r_{p}^{2}\right)^{-2} \;,
 \label{GDipole}
\end{equation}
where $r_{p}^{2}\!=\!0.707\,{\rm fm}^{2}$ is the mean square proton radius. Note that for the proton charge radius we have adopted the 2018 CODATA recommended value 
of $r_{p}\!=\!0.841\,{\rm fm}$~\cite{CODATA}. For the magnetic form factors we use $\mu_{p}\!=\!2.792\,847\,344\,63(82)$ 
and $\mu_{n}\!=\!-1.913\,042\,76(45)$ for the proton
and neutron magnetic moments (in units of the nuclear magneton), respectively~\cite{CODATA}. 
\\ 
In turn, for the electric neutron form factor we use the Galster parametrization~\cite{Kelly:2002if}:
\begin{equation}
  G_{\rm E}^{\,n}(Q^{2}) = -\left(\frac{Q^{2}r_{n}^{2}/6}{1+Q^{2}/M^{2}}\right)
  G_{\rm D}(Q^{2})  \;,
\label{NeutronFF}
\end{equation}
where $r_{n}^{2}\!=\!-0.116\,{\rm fm}^{2}$ is the mean square radius of the neutron. All electromagnetic form factors 
are expressed exclusively in terms of the experimentally determined mean square radii and magnetic moments.

Although one could invoke more sophisticated parametrizations that may be extended to large values of $Q^{2}$~\cite{Ye:2017gyb}, for the 
modest values of the momentum transfer probed in this work, $Q\sim 78-292\, \mathrm{MeV}$, 
the two parametrizations are largely indistinguishable;
in the relevant range the differences for $G_{\rm E}^{\,p}(Q^{2})$ amount to less than 0.2\%. {We have also tested that the errors in the form factor parameters are negligible for our purposes}.

We conclude this section by connecting the single nucleon electromagnetic form factors to the corresponding weak form factors. 
Invoking isospin symmetry and neglecting any strangeness contribution\,\cite{Horowitz:2012we} we obtain 
\begin{subequations}
\begin{eqnarray}    
 \widetilde{G}_{E}^p(Q^{2})&=&g_{V}^{p} G_E^p(Q^{2})+g_{V}^{n} G_E^n(Q^{2}),\\
 \widetilde{G}_{E}^n(Q^{2})&=&g_{V}^{n} G_E^p(Q^{2})+g_{V}^{p} G_E^n(Q^{2}), \\
 \widetilde{G}_{M}^p(Q^{2})&=&g_{V}^{p} G_M^p(Q^{2})+g_{V}^{n} G_M^n(Q^{2}),\\
 \widetilde{G}_{M}^n(Q^{2})&=&g_{V}^{n} G_M^p(Q^{2})+g_{V}^{p} G_M^n(Q^{2}),
\end{eqnarray}
\end{subequations}
where $g_{V}^{p}\!=\!0.0709(2)$ and $g_{V}^{n}\!=\!-0.9901(2)$ are the proton and neutron weak vector charges, respectively, including
radiative corrections~\cite{ParticleDataGroup:2022pth}.

\subsection{Spin-Orbit Currents}

The importance of the so-called spin-orbit currents was {studied} in Ref.~\cite{Ong:2010gf} 
and shortly after extended to the electroweak sector~\cite{Horowitz:2012we}. The nucleon 
matrix elements of both the electromagnetic current as well as the vector part of the weak 
neutral current may in general be written in terms of Lorentz vector and tensor components 
multiplied by Dirac and Pauli (single-nucleon) form factors, respectively. These single-nucleon 
form factors encode the quark substructure of the nucleon. In particular, the spin-orbit 
contribution to the electroweak current is associated with the tensor component of the nucleon 
electromagnetic current.

Although in general small and often neglected, the spin-orbit current makes a significant 
contribution to the form factor when one of the spin-orbit partner states is not fully occupied. 
Indeed, in the CREX analysis\,\cite{CREX:2022kgg}, it was critical to include the spin-orbit 
contribution given that in ${}^{48}$Ca the neutron $f_{7/2}$ orbital is full while its $f_{5/2}$ 
spin-orbit partner is empty. This may suggest that for ${}^{12}$C, with both proton and neutron 
$p_{3/2}$ orbitals filled but its spin-orbit partner $p_{1/2}$ empty, spin-orbit currents may 
play an important role. However, as we demonstrate in Appendix \ref{App:spin-orbit}, this is not 
the case: its contribution to both $R_{\rm ch}$ and $R_{\rm wk}$ does not exceed 0.05\%.

\subsection{Nuclear Radii }

Using $R_{\rm ch}^{\rm SOG}\!=\!2.469(5)\,\mathrm{fm}$ and neglecting spin-orbit effects (0.05\%) in Eq.~\eqref{Rch2}, one extracts an ``experimental" point-proton radius of $R_{p}=2.389(5)\,{\rm fm}$.
There have been several attempts to extract the point-neutron radius of ${}^{12}$C using hadronic probes. For example, 
an elastic proton scattering experiment with ${}^{12}$C carried out at the Los Alamos Meson Physics Facility (LAMPF) 
reported $R_{n}\!=\!2.4\pm\!0.1\,{\rm fm}$\,\cite{Blanpied:1978ta}, a central value that is surprisingly 
larger than the proton one, albeit with a large error. More recently, interaction, reaction, and charge-exchange cross 
sections have been used to extract both point-proton and matter radii\,\cite{Ozawa:2001hb,Kanungo:2016tmz}, from which 
one extracts a neutron radius of $R_{n}\!=\!2.35(4)\,{\rm fm}$. Based on these results, one must conclude that the precise 
extraction of the neutron radius of ${}^{12}$C using hadronic probes is challenging. Perhaps the safest assumption is that 
proton and neutron radii are equal within errors, given that it is a nucleus with $N\!=\!Z$. However, 
the repulsive Coulomb 
interaction pushes the protons out relative to the neutrons suggesting a negative neutron skin. 
Therefore, we now proceed
to estimate the nuclear structure uncertainties associated with the predictions for the point proton and neutron densities.

The ground-state proton and neutron densities for ${}^{12}$C are computed using a large set of nonrelativistic and covariant 
energy density functionals calibrated to the binding energy and charge radii of a collection of magic and semi-magic nuclei 
that do not include ${}^{12}$C. Folding these point-nucleon densities with the single-nucleon form factors one obtains
the charge and weak-charge densities that serve as input for the distorted wave calculation of the parity violating asymmetry. 
All the considered models fall within the general framework of energy density functionals. To avoid any possible bias in the analysis it would be useful to extend the study to ab-initio predictions for the proton and neutron distributions in ${}^{12}$C, which unfortunately are not available for such a light nucleus (see Appendix~\ref{App:EDFvsabinitio} for a tentavite discussion on model consistency).

\section{Analysis}

To assess the nuclear-structure uncertainty, we display in Fig.~\ref{fig:Rwkcorr} model predictions for the weak (vertical axis) 
and charge (horizontal axis) radii of ${}^{12}$C. Also shown in the figure is the experimental charge radius
$R_{\rm ch}^{\rm exp}\!=R_{\rm ch}^{\rm SOG}=\!2.469(5)\,{\rm fm}$~\cite{DeVries:1987atn}. 
As anticipated, given that the major difference between the two radii is due to the Coulomb repulsion, all models 
predict a small negative weak skin of about $R_{\rm wk}\!-\!R_{\rm ch}\!\approx\!-0.015\,{\rm fm}$. 
\begin{figure}[h]
    \centering
    \includegraphics[width=\columnwidth]{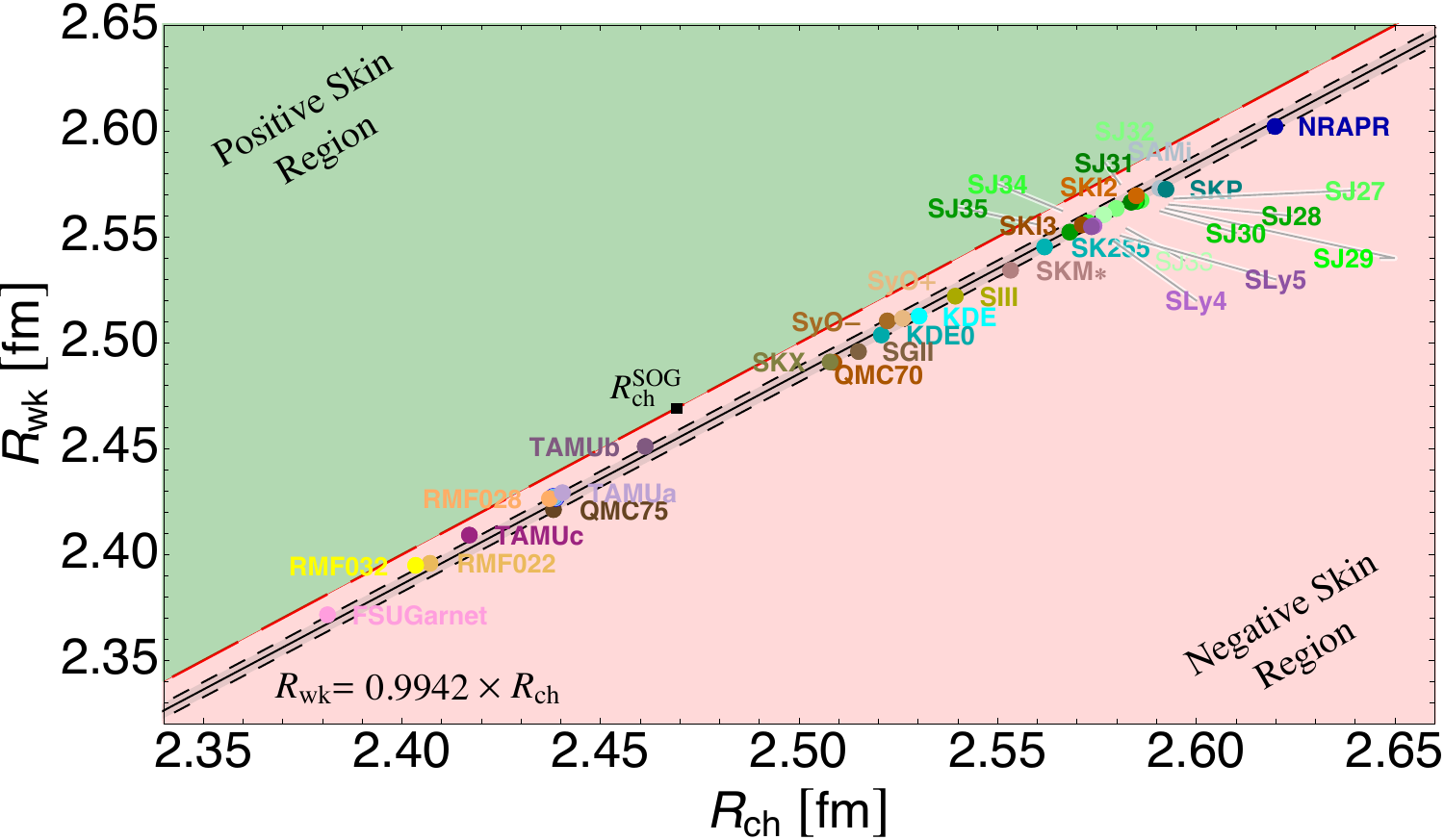}
    \caption{Scatter plot of the charge and weak radii obtained from a variety of nuclear models. The gray band shows the linear 
    fit to the data points, obtained assuming a 1\% uncertainty of the model prediction for the weak radius. The black square is 
    placed at the experimental value of the charge radius, $R_{\rm ch}=2.469(5)$~fm, assuming a zero weak skin.}
    \label{fig:Rwkcorr}
\end{figure}

The new observation in Fig.~\ref{fig:Rwkcorr} is a strong correlation between the model predictions for the
weak and charge radii, generally expected based on the 
strong neutron-proton nuclear interaction.
This strong correlation is shown by the black line in Fig.~\ref{fig:Rwkcorr} which was obtained by performing a linear fit to the model predictions. 
To perform a statistical analysis and define a $\chi^2$ function, we assigned a conservative 1\% uncertainty to each prediction \cite{Chen:2014sca, Roca-Maza2015}. This procedure returns the following
estimate for the weak radius of ${}^{12}$C: 
\begin{equation}
    R_{\rm wk}=0.9942(15)\times R_{\rm ch}\, .\label{RwkvsRchFit}
\end{equation}
The gray band in Fig.~\ref{fig:Rwkcorr} indicates the corresponding $\pm1 \sigma$ region.
%
\begin{figure*}[ht]
    \centering
    \includegraphics[width=1.1\textwidth]{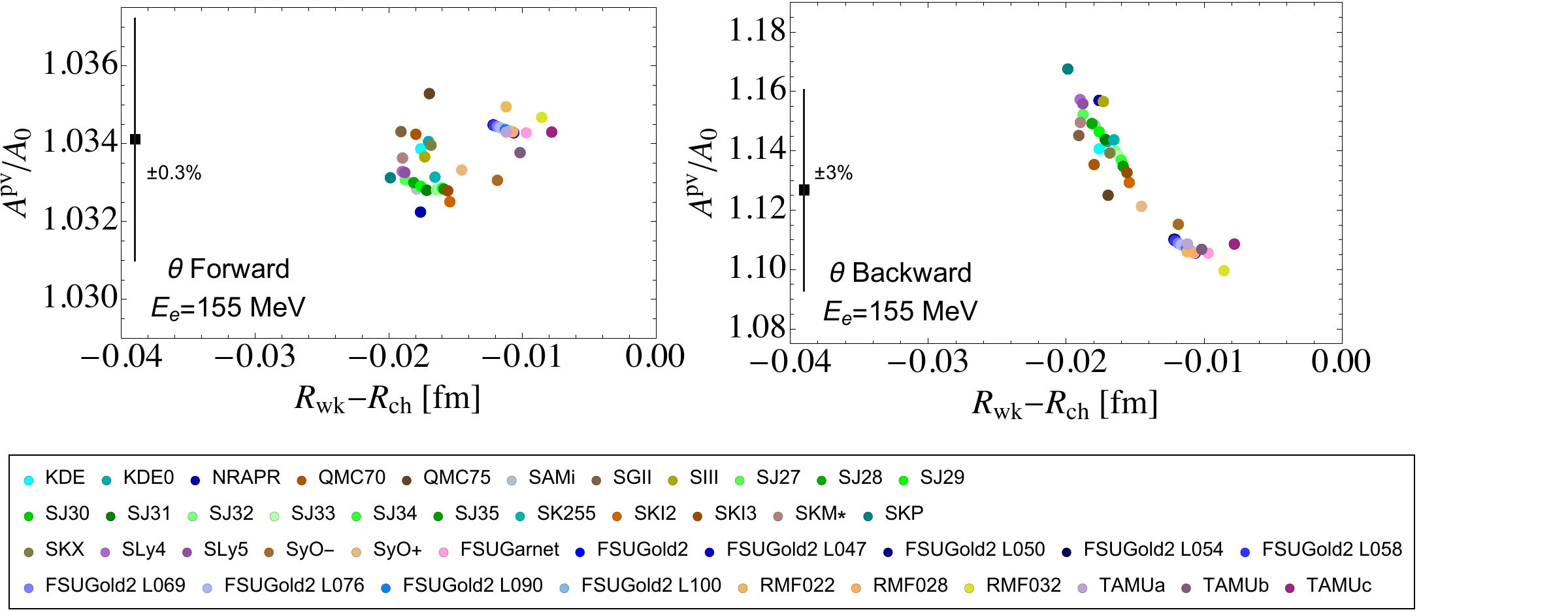}
    \caption{Asymmetry normalized to the pre-factor $A_0$ at the appropriate value of $Q^2$, evaluated for all non-relativistic \cite{KDE, NRAPR, QMC, BSK17, SKX, SIII, SGII, SKM*, SKP, SKI, SLY, SKO, SAMI, SAMIJ, SK255} and covariant \cite{Chen:2014sca,Chen:2014mza,Fattoyev:2013yaa} models using only the nuclear model densities. Each point is set at the corresponding prediction of the weak skin. The left plot shows the results for forward kinematics ($\theta_{\rm f}=29^{\circ}$), the right plot for backward kinematics ($\theta_{\rm b}=145^{\circ}$). The black squared data points with error bars at the left border show the expected precision for the asymmetry measurements centered at an arbitrary value.}
    \label{fig:Fforwardbackward}
\end{figure*}

It is informative to 
note a discrepancy between two sets of models: while all the Skyrme models systematically over-predict the charge radius, 
the relativistic models do the opposite. The source for such a discrepancy may have its origin in the competition between the 
saturation density and the surface energy. Skyrme interactions tend to predict that symmetric nuclear matter saturates at a 
higher density than the covariant models, typically $0.16\,{\rm fm}^{-3}$ versus $0.15\,{\rm fm}^{-3}$. On its own, this would
suggest that Skyrme EDFs predict smaller nuclear radii than covariant EDFs. However, both sets of functionals use charge radii 
of medium to heavy nuclei as input for the calibration. This indicates that, in order for both sets of functionals to reproduce the 
same charge radius of, e.g., ${}^{208}$Pb, the surface energy in the Skyrme models must be softer than in the relativistic models. 
However, ${}^{12}$C is relatively light to display any evidence of a saturated core, so its charge radius is dominated by the surface 
energy---leading to systematically larger radii for Skyrme EDFs than for covariant EDFs.

Returning to the discussion of Fig.~\ref{fig:Rwkcorr}, we observe that most of the models fail to reproduce the experimental value of the charge radius $R_{\rm ch}^{\rm exp}$ by as much as $\pm 5\%$. 
To circumvent this significant model dependence, in Ref.~\cite{Koshchii:2020qkr}, some of the authors of the present work hand-picked just five nuclear mean-field models by requiring that the charge radius be predicted within $\approx1\%$ from the experimental one. Consequently, the respective model predictions for the weak density (fitted to a 2-parameter Fermi function form\,\cite{Piekarewicz:2016vbn} to accelerate the numerical evaluation) were used to quantify the nuclear theory uncertainty for the weak skin extracted from a possible future measurement of the PV asymmetry. In doing so, the charge density was fixed to the experimental one (i.e., SOG). The conclusion of 
Ref.~\cite{Koshchii:2020qkr} was that the extraction of the weak charge of $^{12}$C from a 0.3\% measurement of the PV asymmetry at low momentum transfer was contaminated by the nuclear theory uncertainty, such that an additional measurement at a larger value of the momentum transfer was necessary.

Here we argue that the strong correlation between the weak and charge radii observed in Fig.~\ref{fig:Rwkcorr}
makes this arbitrary model selection  unnecessary, allowing for a much more reliable statistical analysis on a large sample of nuclear models. 
Therefore, for consistency and accuracy, we prefer to directly use the theoretical density predicted by the nuclear models instead of a two-parameter Fermi function fitted to it. 
Finally, we take into account that nuclear models predict point-proton and point-neutron densities, rather than the experimentally measured weak and charge densities. Hence, to prevent the introduction of additional systematic errors, we determine the weak and charge densities by folding the theoretical predictions with the single-nucleon form factors introduced in Sec.\ref{sec:nucleonFF}.\\
To further illustrate the importance of maintaining the $R_{\rm wk}\!-\!R_{\rm ch}$ correlation, we display predictions for the PV asymmetry in Fig.~\ref{fig:Fforwardbackward}. As mentioned earlier, for a realistic prediction for the PV asymmetry one has to go beyond PWBA and account for Coulomb distortions. This entails numerically solving the Dirac equation for an electron moving in the Coulomb field of the nuclear charge distribution supplemented by a PV weak interaction potential which is the source of the asymmetry.
As such, the main input into the DWBA code are the nuclear weak and charge densities which we consistently take from each of the models adopted in this work.

As in Ref.~\cite{Koshchii:2020qkr}, we explore two scenarios at a fixed electron beam energy of $E_e=155\ \mathrm{MeV}$: A forward ($\theta_{\rm f}=29^{\circ}$) and backward 
($\theta_{\rm b}=145^{\circ}$) angle configuration to highlight two different features of the PV asymmetry.
Results for the PV asymmetry are displayed in Fig.~\ref{fig:Fforwardbackward}; left for the forward angle measurement and right for the backward 
one. We find that in this new approach the spread of both the weak skin and the PV asymmetry is dramatically reduced relative to the 
uncertainty found in Ref.~\cite{Koshchii:2020qkr}. 

We observe on the left-hand panel of Fig.~\ref{fig:Fforwardbackward} that the predictions for the PV asymmetry at the forward angle are 
fairly scattered without following a particular trend as a function of the weak skin. However, whereas the spread in the weak skin is relatively 
large for such a light nucleus, it does not lead to any significant spread in $A^{\rm PV}$. Indeed, the range of the predicted PV
asymmetry for the forward measurement is within the $0.3\%$ precision goal, as indicated by the vertical error bar with the central value (black
squared point) placed arbitrarily around 1.034. {We note that a measured central value differing from 1.034 by more than a 0.3\% would clearly imply a flaw in current nuclear structure models based on Density Functional Theory. The latter would be a very interesting although very unexpected result.}

The right-hand panel of Fig.\,\ref{fig:Fforwardbackward} contains the same information, but this time for a backward scattering angle. The models display a systematic trend indicating that the PV asymmetry is anti-correlated to the weak skin. Although we have argued on the importance of Coulomb distortions, the PWBA provides a simple explanation for this anti-correlation. Expanding both form factors in Eq.~\eqref{ApvBorn} to first order in $Q^{2}$ one obtains 
\begin{equation}
   \frac{A^{\rm PV}_{\rm PWBA}}{A_{0}} \simeq
   1 - \frac{Q^{2}}{3}R_{\rm ch}R_{\rm wskin}
\end{equation}
where $R_{\rm wskin}\!=\!R_{\rm wk}\!-\!R_{\rm ch}$ is the weak skin. 
\\

Armed with these observations, we proceed to an analysis of the statistical ensemble of model predictions for  $A^{\rm PV}$ in the two experimental scenarios, for forward and backward scattering angles. 

We start with the forward measurement and notice that while the correlation between $A^{\rm PV}$ and the weak skin in Fig.~\ref{fig:Fforwardbackward} is weak, a stronger correlation emerges if the asymmetry is plotted versus the charge radius, as shown in Fig.~\ref{fig:Apv_corr_F}.
In this figure, the black solid line indicates the result of a linear fit. 
To define the $\chi^2$ of this fit, we assign  to each individual model prediction a conservative 1\% uncertainty.
The gray band in the figure indicates the resulting $1\sigma$ region obtained in this manner. The vertical red line is set to the experimental value 
of the charge radius of $^{12}$C. While its uncertainty is $\sim0.2\%$, even an overly conservative $\pm2\%$ range 
translates into a corresponding uncertainty for the PV asymmetry of about $0.18\%$, as indicated by the orange box and lines. Given that
this theoretical uncertainty is below the anticipated precision goal of the experiment, we conclude that the determination of the weak charge 
of $^{12}$C from the forward measurement alone could be considered a clean measurement of the weak mixing angle, without any significant 
contamination from nuclear structure effects.

\begin{figure}[h]
    \centering
    \includegraphics[width=\columnwidth]{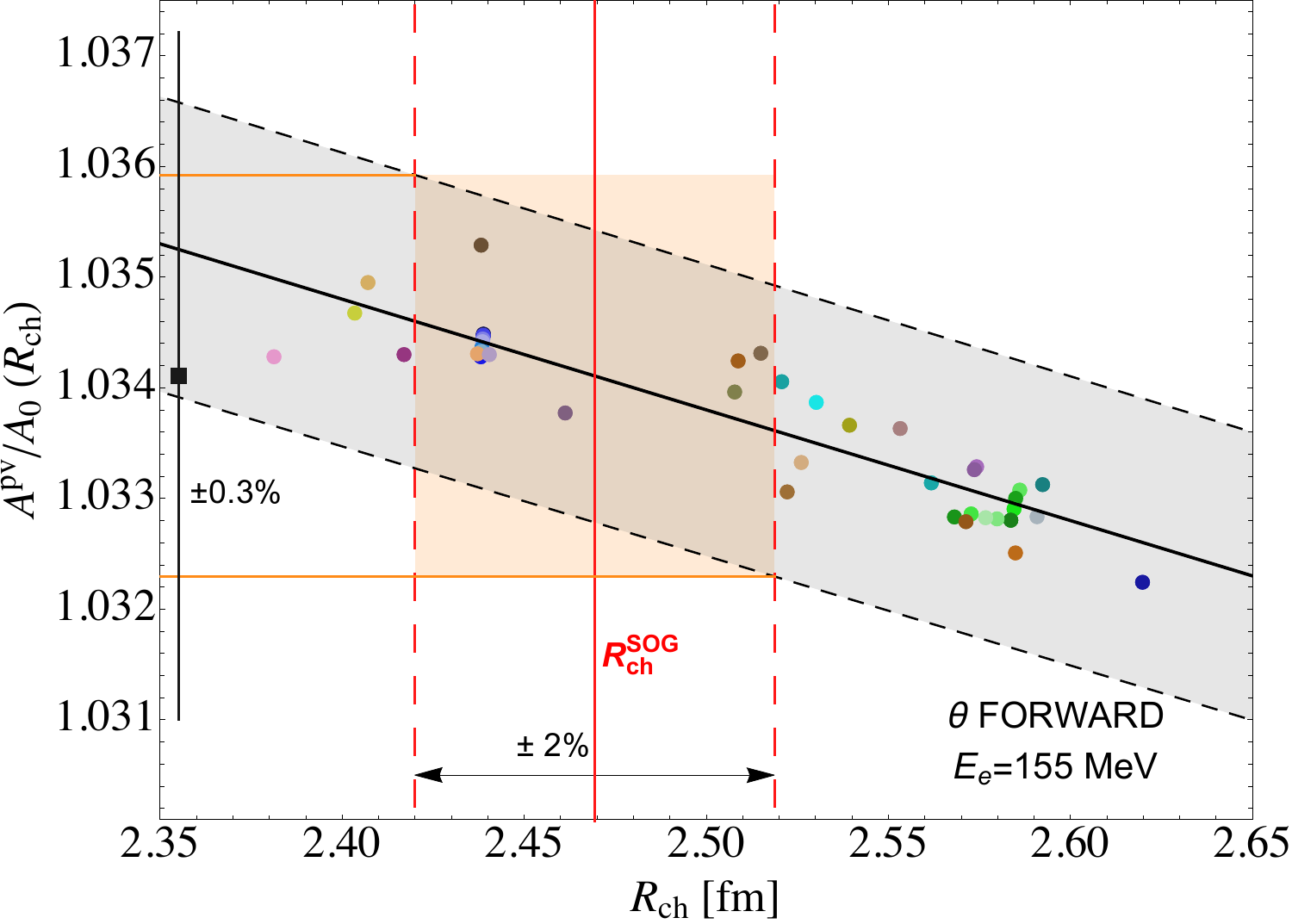}
    \caption{Linear fit to the predictions of the forward asymmetry normalized to the pre-factor $A_0$ (black solid lines) as a function of the charge radius. The $1\sigma$ region is shown by the gray band. The red solid line shows the measured value of the charge radius, while the dashed vertical lines indicate its $\pm2\%$ uncertainty region. The orange box indicates the estimated uncertainty on the asymmetry from the correlation of the nuclear models.}
    \label{fig:Apv_corr_F}
\end{figure}

We next turn to the analysis of the backward measurement of $A^{\rm PV}$. Its observed sensitivity to the weak skin motivates us to perform a linear fit of the following form:
\begin{equation}
A^{\rm PV}_b(R_{\rm wskin})=\mathcal{A} R_{\rm wskin} +\mathcal{B}\, . 
\label{eq:BackPred}
\end{equation}
To perform the linear fit, we again assign to each individual $A^{\rm PV}$ prediction an uncertainty of $1\%$. 
\\ 
The linear fit returns 
\begin{eqnarray}\label{AandBbestFit}
    \mathcal{A}&\!=\!&(-39\pm 4)\times10^{-6}\, \mathrm{fm}^{-1}\, ,\\\nonumber
    \mathcal{B}&\!=\!&(7.37\pm 0.05)\times10^{-6} 
\end{eqnarray} 
for the two coefficients, and is shown in Fig.~\ref{fig:Apv_corr_B} by a black line with the shaded gray band indicating the $1\sigma$ region. 
Different experimental scenarios for the attainable precision of the backward measurement, denoted by  
$\epsilon_b$, are indicated by colored horizontal bands: an orange band for $\epsilon_b\!=\!3\%$, green for $\epsilon_b\!=\!5\%$, and blue for $\epsilon_b\!=\!7\%$. 
Since the backward measurement retains sensitivity to both the weak charge and weak skin, we must resort to a combined analysis of the forward and backward angle results, which we will discuss in the next section.
\begin{figure}[h]
    \centering
    \includegraphics[width=\columnwidth]{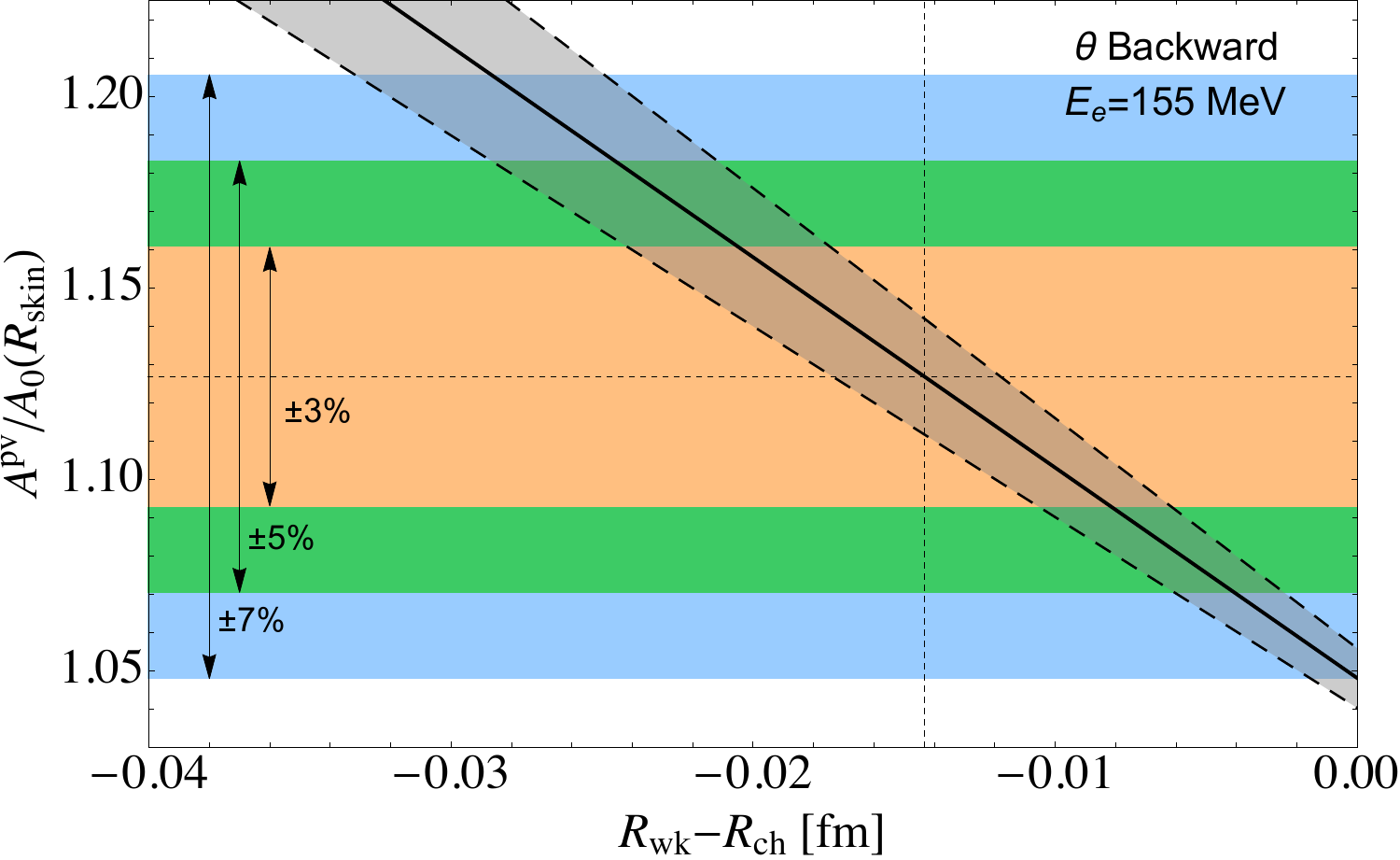}
    \caption{Linear fit to predictions of the backward asymmetry normalized to the pre-factor $A_0$ as a function of the weak skin (black solid line). The $\pm1\sigma$ uncertainty range is shown by the gray band. The short-dashed lines indicate reference values for the weak skin and the corresponding asymmetry. The colored horizontal bands indicate the different precision scenarios for a possible measurement: $\epsilon_b=3\%$ (orange), $\epsilon_b=5\%$ (green) and $\epsilon_b=7\%$ (blue).}
    \label{fig:Apv_corr_B}
\end{figure}

\section{Combined sensitivity analysis \\ 
for the weak mixing angle \\ 
and the weak skin}
In this section we explore the impact of a combined analysis of the forward and backward measurements of 
the PV asymmetry on the simultaneous determination of the weak mixing angle and the weak skin of $^{12}$C. 
We have established already that the forward measurement at the $0.3\%$ precision (or worse) provides a clean determination of the nuclear weak 
charge. 
The backward measurement retains  sensitivity to both the weak charge and weak 
skin. \\

To analyze the backward measurement, we 
take as reference values $R^{\rm ref}_{\rm wk}\!=\!2.455\,\mathrm{fm}$ for the weak radius, which has been obtained by exploiting the correlation in Eq.~\eqref{RwkvsRchFit} and $R_{\rm ch}=R_{\rm ch}^{\rm SOG}$. The latter weak radius corresponds to a weak skin of
$R^{\rm ref}_{\rm wskin}\!=\!-0.0143\,\mathrm{fm}$, which falls about in the middle of the predictions of the considered models sample. These choices imply a corresponding 
reference value for the PV asymmetry at backward angles of $A^{\rm PV,ref}_{b}\!=\!7.92$\,ppm, obtained by using the definition in Eq.~\eqref{eq:BackPred} (and shown in Fig.~\ref{fig:Apv_corr_B} by the dashed black horizontal line). 
As the asymmetry at forward angles is practically insensitive to the reference value chosen for the weak skin, we consider as a reference prediction an average value obtained from the considered nuclear models, which numerically corresponds to $A^{\rm PV,ref}_{f}\!=\!0.517$\,ppm (and is shown by the black squared data point in Fig.~\ref{fig:Apv_corr_F}).
The optimization of the parameters is obtained by minimizing the following $\chi^2$-function:
\begin{align}
   \chi^2&=\left(\dfrac{A_{f}^{\rm PV,ref}-\eta_1 A^{\rm PV}_{f}(\sin^2\theta_W)}{\epsilon_f} \right)^2\nonumber\\
    &+\left(\dfrac{A_{b}^{\rm PV,ref}-A^{\rm PV}_b(\eta_2,\sin^2\theta_W,R_{\rm wk})}{\sigma_b} \right)^2\nonumber\\
   &+\left(\dfrac{\eta_1-1}{\sigma_{\eta_1}}\right)^2+\eta_2^2\, ,\label{ChiSquareFinale}
\end{align} 
which combines the information from the forward angle measurement with the fixed precision goal of $\epsilon_f=0.3\%$ (first term),
information from the backward measurement with different precision scenarios for $\epsilon_b$ (second term), and information on 
the model uncertainty (last terms). The adopted uncertainty in $\sigma_{b}$ is obtained by adding in quadratures the anticipated
experimental error $\epsilon_b$ with the uncertainty in the $\mathcal{B}$ parameter associated with the linear fit given in Eq.~\eqref{AandBbestFit}, $\sigma_b^2=\epsilon_b^2+\sigma_{\mathcal{B}}^2$.
By writing the asymmetry as a function $A_{f}^{\rm PV}(\sin^2\theta_W)$, we explicitly indicate its dependence on the weak mixing angle from its relation to the nuclear weak charge, see Eq.~\eqref{ApvBorn}.
Further, $\eta_1$ and 
$\sigma_{\eta_1}$ are introduced as nuisance parameters to account for our estimate of the nuclear uncertainty, i.e.\ 
$\sigma_{\eta_1}=0.18\%$. This allows the PV asymmetry at the forward angle to vary withing the orange box indicated in Fig.~\ref{fig:Apv_corr_F}. Finally, to describe the uncertainty of the parameter $\mathcal{A}$ in Eq.~\eqref{AandBbestFit} we introduce the nuisance parameter $\eta_2$ and parametrize the backward asymmetry as follows:
\begin{align}
    A^{\rm PV}_b&(\eta_2,\sin^2\theta_W,R_{\rm wk})=\dfrac{Q_W(\sin^2\theta_W)}{Q_W^{\rm SM}} 
    \nonumber \\
    \times&\left[\left(\mathcal{A}+\eta_2\,\sigma_{\mathcal{\!A}}\right)R_{\rm wskin}+\mathcal{B}\right]\, .
    \label{APVbparam}
\end{align}
The dependence on the weak mixing angle is restored by rescaling the weak charge as in the case of the forward asymmetry. 
The second line in Eq.~\eqref{APVbparam} is introduced to describe the dependence on the nuclear models, as shown in Fig.~\ref{fig:Apv_corr_B} 
and Eq.~\eqref{eq:BackPred}. The quantity 
$\sigma_\mathcal{A}$ is the uncertainty of 
$\mathcal{A}$, as reported in Eq.~\eqref{AandBbestFit}.  
$\eta_2$ is defined such that for $\eta_2=\pm1$ one retrieves the $\pm 1\sigma$ value for $\mathcal{A}$, i.e.\ the slope of the prediction for the asymmetry goes from the lower to the higher dashed black lines in Fig.~\ref{fig:Apv_corr_B}. 

\begin{figure}[h]
    \includegraphics[width=\columnwidth]{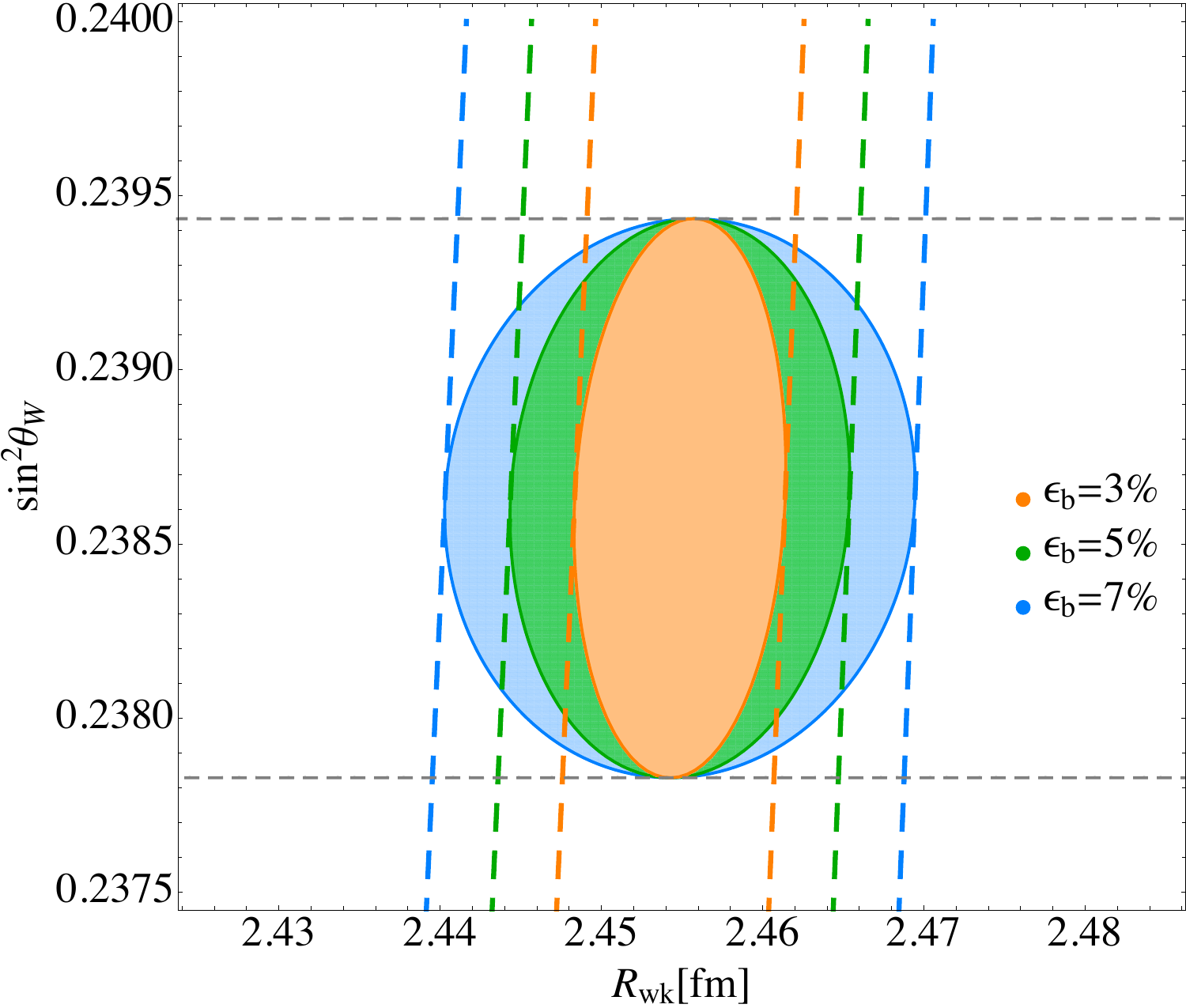}
    \caption{Results of the sensitivity to the weak mixing angle and the weak radius for the combined forward-backward measurement. The different $\Delta\chi^2=1$ contours correspond to the different precision scenarios for the backward measurement (orange for $\epsilon_b=3\%$, green for $\epsilon_b=5\%$ and blue for $\epsilon_b=7\%$). The dashed lines indicate the results of the independent analyses of the forward and backward measurements.}
    \label{fig:AvsRchorRwkB}
\end{figure}
 
The results of the combined fit to both the forward and backward measurements are displayed in Fig.~\ref{fig:AvsRchorRwkB}. The various
contours, obtained by imposing the condition $\Delta\chi^2\!=\!1$, correspond to the three different values for $\epsilon_b$. 

We can extract the individual sensitivities of a combination of the forward and backward measurements to the weak mixing angle and the weak radius by splitting the $\chi^2$-function in Eq.~\eqref{ChiSquareFinale} and retaining only the relevant contribution. 
\\
The horizontal dashed lines show the marginalization in terms of the weak mixing angle: as already mentioned, its extraction is not contaminated by the nuclear model dependence, nor does it depend on the precision of the backward measurement.
Retaining only the information from the backward measurement results in nearly vertical bands in the parameter space. 
\\
The various colors for the dashed vertical bands correspond to different choices for the precision of the backward measurement $\epsilon_b$, as indicated in the legend of the figure. The fact that these bands are slightly tilted indicates the residual (mild) dependence on the weak mixing angle.
Therefore, a combined analysis of the forward and backward measurements is necessary to simultaneously extract both parameters.
\begin{table}[h]
\resizebox{\columnwidth}{!}{
\renewcommand{\arraystretch}{1.8}
\begin{tabular}{|c|c|c|c|c|}
\hline
$\epsilon_f$ &  $\epsilon_b$ & $R_{\rm wk}$ $[{\rm fm}]$ & $\sigma_{R_{\rm wk}}$ $[\%]$ & $\sin^2\theta_W$ \\ \hline
\multirow{3}{*}{$0.3\%$}      & $3\%$    & $2.455\pm0.008$     &   $0.34\%$     & $0.2386\pm0.0008$                                     \\ \cline{2-5} 
     & $5\%$        & $2.455\pm0.012$   &   $0.48\%$    & $0.2386\pm0.0008$                                     \\ \cline{2-5} 
     & $7\%$         & $2.455\pm0.016$    &   $0.63\%$    & $0.2386\pm0.0008$                                      \\ \hline
\end{tabular}}
\caption{Summary of the best fit and precision values  corresponding to the results shown in Fig.~\ref{fig:AvsRchorRwkB}. 
\label{TableResults}
}
\end{table}

The numerical results obtained by marginalizing the contours shown in Fig.~\ref{fig:AvsRchorRwkB} are summarized in Tab.~\ref{TableResults}, where we collect the best fit values and the relative uncertainties of the weak radius and the weak mixing angle. The assumed precision of the forward measurement $\epsilon_f=\!0.3\%$ directly translates into that for the weak mixing angle. Similarly, depending on the precision assumed for the backward measurement, the precision of the weak radius ranges between $\sim\!0.34\%$ and $\sim\!0.63\%$. 

We have also explored a wider set of values for the precision of a backward measurement. Results are shown in Fig.~\ref{fig:ptPrecision}, where $\epsilon_b$ is varied between 1\% and 8\%. The figure shows the achievable precision for the weak radius (both relative and absolute).
\begin{figure}[h]
    \includegraphics[width=\columnwidth]{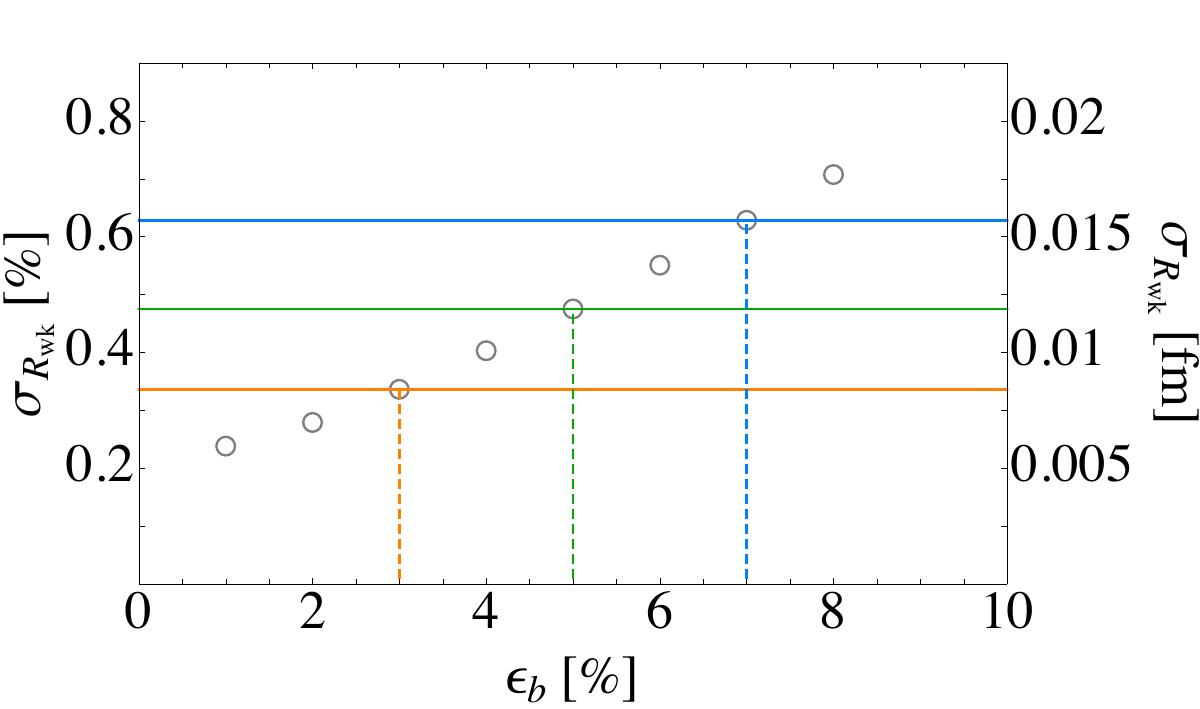}
    \caption{Precision for the weak radius as a function of the precision of the backward asymmetry. The colored lines indicate the precision scenarios considered in Ref.~\cite{Koshchii:2020qkr}.}
    \label{fig:ptPrecision}
\end{figure}

The impact of a sub-percent extraction of the weak mixing angle at low energies is demonstrated in Fig.~\ref{fig:s2wRunning} where we display $\sin^2\theta_\mathrm{W}$ along with existing measurements~\cite{ParticleDataGroup:2024pth}.\\ 
One can see that a $0.3\%$ measurement of the PV asymmetry with $^{12}$C at MESA has the potential of providing one of the most precise measurements of 
the weak mixing angle at low momentum transfers. This could allow one to obtain new limits on models for physics beyond the SM. As an example 
we show in Fig.~\ref{fig:s2wRunning} how the running of the weak mixing angle could be modified if an additional light vector mediator, 
such as a dark $Z$ boson~\cite{Davoudiasl:2012ag,Davoudiasl:2012ig,Davoudiasl:2013aya,Davoudiasl:2012qa,Arcadi:2019uif,Cadeddu:2021dqx},  existed. 
In such a model, the effect on the running is significant only at low scales, depending on the coupling and the mass of the new boson. The dashed 
blue lines shown in Fig.~\ref{fig:s2wRunning} correspond to a model with a $\sim\!50\, \mathrm{MeV}$ dark $Z$ boson, using the results obtained by 
the analysis in Ref.~\cite{Cadeddu:2021dqx}. Having a sub-percent measurement of the weak mixing angle at around $Q\sim\!78\, \mathrm{MeV}$ will help constraining a hypothetical deviation of the running weak mixing angle from the SM prediction. 
\begin{figure}[h]
    \centering
    \includegraphics[width=\columnwidth]{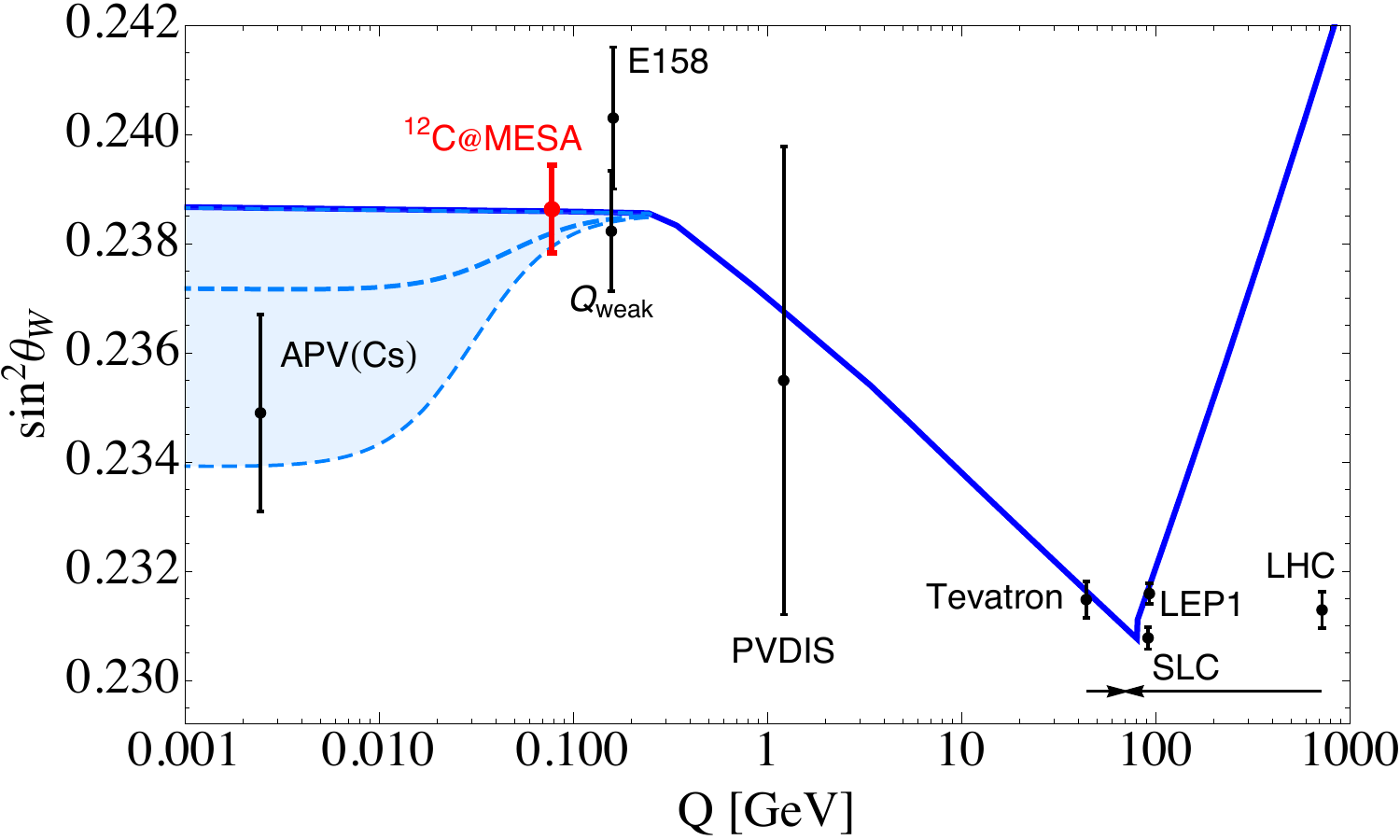}
    \caption{Current status of $\sin^2\theta_W$ measurements as a function of the energy scale $Q$ ~\cite{Wood:1997zq,Dzuba:2012kx,Qweak:2021ijt,SLACE158:2005uay,PVDIS:2014cmd,ParticleDataGroup:2024pth}. The blue solid curve represents the SM prediction for the running weak mixing angle~\cite{ParticleDataGroup:2022pth}. The red data point shows our result and is set at $Q\sim 78\ \mathrm{MeV}$. The light blue dashed curves and the blue filled region indicate the modification of the running weak mixing angle in the presence of a dark $Z$ boson, as discussed in Ref.~\cite{Cadeddu:2021dqx}.}
    \label{fig:s2wRunning}
\end{figure}

As a final note we point out that the possibility to cleanly determine the point-neutron radius and the neutron skin of the stable daughter nuclei participating in the superallowed nuclear $\beta$ decays will have an impact on the extraction of the Cabibbo-Kobayashi-Maskawa quark-mixing matrix element $V_{ud}$~\cite{Seng:2022epj,Gorchtein:2023naa}. Future measurements of the PV asymmetry in PVES with nuclei such as $^{26}$Mg, $^{42}$Ca or $^{54}$Fe can be carried out at the MESA facility. A dedicated feasibility study for each of these nuclei, similar to the $^{12}$C primer performed here, will be necessary.

\section{conclusions}
In conclusion, we have revisited the physics reach of a measurement of the parity violating asymmetry for   
$^{12}\mathrm{C}$. Improvements with respect to the previous work reported in Ref.~\cite{Koshchii:2020qkr} include a much larger sample of
nuclear models analyzed, and
a more robust procedure of assessing the nuclear theory uncertainties. 
We observed a tight correlation between the nuclear models' predictions for the weak and charge radii. The difference of the two, the weak skin, is expected to be small and negative due to the Coulomb repulsion among the protons inside an $N\!=\!Z$ nucleus.
Predictions for the asymmetry at forward angles show no significant dependence on the weak skin, and solely remain correlated with the charge radius which is known experimentally with a sufficient precision. Thus, the forward measurement of the PV assymetry of $^{12}$C planned at MESA will determine the weak mixing angle at low momentum transfer, not contaminated by nuclear uncertainties.
The asymmetry in the backward kinematics is, instead, strongly correlated with the weak skin. This correlation has 
been exploited to assess the precision required to provide a clean electroweak determination of the weak radius of ${}^{12}$C. We performed 
a combined analysis of the forward-backward measurement and showed
that assuming the planned $0.3\%$ precision at forward angles, it directly translates into a $0.3\%$ determination of the weak mixing angle; meanwhile, a $3\%$ measurement at backward angle will 
result in the first electroweak determination of the weak radius of $^{12}\mathrm{C}$ with a precision of $\sim\!0.34\%$. More aggressive scenarios naturally result in an even more precise determination of the weak radius. 
We discussed the consequences of such an ambitious experimental program for precision tests of the standard model and beyond with the running weak mixing angle.

\begin{acknowledgments}
XRM acknowledges support by grants PID2020-118758GB-I00 funded by MCIN/AEI/10.13039/5011 00011033; by the ``Unit of Excellence Mar\'ia de Maeztu 2020-2023'' award to the Institute of Cosmos Sciences, Grant CEX2019-000918-M funded by MCIN/AEI/10.13039/501100011033; and by the Generalitat de Catalunya, grant 2021SGR01095. JP acknowledges support from the U.S. Department of Energy Office of Science, Office of Nuclear Physics under Award Number DE-FG02-92ER40750. M.G. acknowledges support by EU Horizon 2020 research and innovation programme, STRONG-2020 project
under grant agreement No 824093, and by the Deutsche Forschungsgemeinschaft (DFG) under the grant agreement GO 2604/3-1.

\end{acknowledgments}

\appendix
\section{Nuclear form factors for non-relativistic energy density functionals}
\label{app:corr}
In Sec.~\ref{sec:nucleonFF} we describe the formalism adopted to treat the nucleon form factors. Those are at the basis of the construction of the nuclear form factors. Here, we report the procedure employed in order to obtain the charge and weak form factors for non-relativistic energy density functionals, which follows closely Refs.~\cite{Horowitz:2012we,Reinhard:2021gym}.
The charge density is calculated as
\begin{equation}
    \rho_{\rm ch}(\mathbf{r})=\dfrac{1}{(2\pi)^3}\int d^3q e^{-i \mathbf{q}\cdot\mathbf{r}} F_{\rm ch}(\mathbf{q}),
\end{equation}
and the charge form factor as~\cite{Reinhard:2021gym}
\begin{eqnarray}
    F_{\rm ch}(\mathbf{q})&=&\sum_{t=p,n}\Big[G^t_{E}(\mathbf{q})\Big(1-\frac{1}{2}\mathbf{q}^2 \mathcal{D}\Big)F_t(\mathbf{q})\nonumber\\
    &-&\mathcal{D}\Big(2 G_{M}^t(\mathbf{q})-G_{E}^t(\mathbf{q})\Big)F_{ls}^t(\mathbf{q})\Big],\label{eq:Fchdef}
\end{eqnarray}
with $\mathcal{D}=\dfrac{\hbar^2}{(2mc)^2}$,  and~\cite{Reinhard:2021gym}
\begin{eqnarray}
    F_t(\mathbf{q})&=&\int d^3r e^{i \mathbf{q \cdot r}}\rho_t(\mathbf{r})\ ,\\
    F_{ls,t}(\mathbf{q})&=&\int d^3r e^{i \mathbf{q \cdot r}}\nabla\cdot\mathbf{J}_t(\mathbf{r})\ .\label{Flst}
\end{eqnarray}
$F_t(\mathbf{q})$ is the point-proton (neutron) form factor, while $F_{ls,t}(\mathbf{q})$ is a form factor related to the current density.
\\
The weak charge density is the Fourier transform of the weak form factor which is introduced in full analogy to Eq.~\eqref{eq:Fchdef}, 
\begin{align}
    F_{\rm wk}(\mathbf{q})&=\sum_{t=p,n}\Big[\widetilde{G}^t_{E}(\mathbf{q})\Big(1-\frac{1}{2}\mathbf{q}^2 \mathcal{D}\Big)F_t(\mathbf{q})\nonumber\\
    &-\mathcal{D}\Big(2 \widetilde{G}_{M}^t(\mathbf{q})-\widetilde{G}_{E}^t(\mathbf{q})\Big)F_{ls}^t(\mathbf{q})\Big].\label{eq:Fwkdef}
\end{align}
The charge and weak radii are obtained from the respective charge densities as
\begin{eqnarray}
    R^2_{ch}&=&\dfrac{1}{Z}\int r^2 \rho_{ch}(r)d^3r\ , \\
    R^2_{\rm wk}&=&\dfrac{1}{Q_{\rm wk}}\int r^2 \rho_{\rm wk}(r)d^3r\ .
\end{eqnarray}

\section{Spin-orbit currents} 
\label{App:spin-orbit}
Given that the treatment of spin-orbit currents is different for relativistic and 
non-relativistic EDFs, this could potentially introduce an 
additional theoretical uncertainty. Here, we demonstrate that this is not the case. 

The impact of the spin-orbit current on the charge radius of a nucleus with $N\!=\!Z$ was shown to take the following simple form\,\cite{Horowitz:2012we}:
\begin{equation}
  R_{\rm ch}^{2} = R_{p}^2 +  \Big(r_{p}^2 + r_{n}^2 \Big) 
                + \Big(\langle r_{p}^{2}\rangle_{\rm so}+
                           \langle r_{n}^{2}\rangle_{\rm so}\Big),
 \label{Rch2}
\end{equation}
where in general, the spin-orbit contributions to $R_{\rm ch}$ can not be evaluated in closed form. Yet, in Ref.~\cite{Horowitz:2012we}
a good estimate was obtained by assuming that the lower component of the Dirac spinor may be derived from the upper component 
assuming a free space relation. In this case one obtains
\begin{equation}
  \langle r_{p}^{2}\rangle_{\rm so}+\langle r_{n}^{2}\rangle_{\rm so} = \frac{2}{3}\frac{(\kappa_{p}+\kappa_{n})}{M^{2}} \approx -0.004\,{\rm fm}^{2},
 \label{Rch2so}
\end{equation}
where $\kappa_{p}\!=\!+1.793$ and $\kappa_{n}\!=\!-1.913$ are the anomalous magnetic moments of the proton and neutron (in units of the nuclear magneton), respectively. 
Although the magnitude of the contribution from each nucleon is about $4/3M^{2} \approx 0.06\,{\rm fm}^{2}$, their combined contribution 
is more than ten times smaller because of the almost equal but opposite values of the anomalous magnetic moments. 

In the general case where the relation between the upper and lower components is determined dynamically, the effect is enhanced because 
the effective nucleon mass $M^{\star}$ is reduced in the nuclear medium. Yet, the large cancellation induced by the anomalous magnetic 
moments persists, resulting in a combined contribution of 
$\langle r_{p}^{2}\rangle_{\rm so}+\langle r_{n}^{2}\rangle_{\rm so}\!\approx\!-0.007\,{\rm fm}^{2}$. 

Given that the anomalous weak magnetic moments are simply related to the corresponding electromagnetic ones, a similar cancellation occurs 
in the case of the weak charge radius $R_{\rm wk}$. Thus, we conclude that the spin-orbit contribution to both $R_{\rm ch}$ and $R_{\rm wk}$ 
is of the order of 0.05\%.

\section{Comparison between EDF and ab-initio models}
\label{App:EDFvsabinitio}
To test the validity of the sample of models considered in this analysis, which is built within the general framework of energy density functionals, we compared with predictions from ab-initio models.
Although to our knowledge there are no 
ab-initio predictions for the proton and neutron distributions in ${}^{12}$C, predictions are available for the proton and neutron 
radii of ${}^{40}$Ca both in coupled-cluster\,\cite{Hagen:2015yea} and dispersive-optical models\,\cite{Pruitt:2020zbf}. 
Given that ${}^{40}$Ca is a heavier $N\!=\!Z$ nucleus than ${}^{12}$C, one anticipates that the stronger Coulomb repulsion 
in ${}^{40}$Ca will generate a slightly larger negative skin than in ${}^{12}$C. The predictions for the neutron skin thickness of 
${}^{40}$Ca are $-0.0461\!\le\!R_{\rm skin}^{40} ({\rm fm})\!\le\!-0.0400$ for the coupled-cluster model and 
$R_{\rm skin}^{40}\!=\!(-0.051 \pm 0.004)$\,fm for the dispersive-optical model. 
These values compare well with the predictions obtained by the same EDF models employed in this work, namely
$-0.05\!\le\!R_{\rm skin}^{40} ({\rm fm})\!\le\!-0.04$ for non-relativistic models and $-0.053\!\le\!R_{\rm skin}^{40} ({\rm fm})\!\le\!-0.046$ for the covariant ones.
Although the extrapolation from ${}^{40}$Ca to ${}^{12}$C may not be trivial, it is gratifying to see that the set of energy 
density functionals used in this work are entirely consistent with these values for the case of ${}^{40}$Ca, thereby lending credibility to our approach.
Neverthless, new predictions from ab-initio calculations for nuclei as light as ${}^{12}$C would be benefitial for improving the current work in the future.

\newpage
\bibliography{sample}
\end{document}